\documentstyle {article}

\newfont{\ag}{cmsy10.tfm}
\newcommand{\ddo}{differential operator}
\newcommand{\ms}{moduli space}
\newcommand{\ps}{pseudodifferential operator}
\newcommand{\gr}{Grassmannian}
\newcommand{\ei}{eigenvalues}
\newcommand{\ml}{multiplication}

\title{On solutions to the string equation}
\author{Albert Schwarz \\ Department of Mathematics, University of California,
Davis, CA
95616 \\ ASSCHWARZ@UCDAVIS.EDU}

\date{}

\begin{document}

\maketitle
\begin{abstract}
The set of solutions to the string equation $[P,Q]=1$ where $P$ and $Q$ are
\ddo s  is
described.It is shown that there exists one-to-one correspondence
between this set and the set of pairs of commuting \ddo s.This fact permits us
to describe the set of solutions to the string equation in terms of moduli spa-
ces of algebraic curves,however the direct description is much simpler.
 Some results are obtained for the superanalog to the string equation where
 $P$ and $Q$ are considered as super \ddo s. It is proved that this equation
 is invariant with respect to Manin-Radul, Mulase-Rabin and Kac-van de Leur
 KP-hierarchies.
\end{abstract}

   Let us consider ordinary \ddo s

\begin {equation}
P=\sum_{m=1}^p a_m (x)\partial ^m, \ \ \ Q=\sum_{n=1}^h b_n (x) \partial ^n
\end {equation}
where $\partial={\partial \over \partial x},\  a_m(x)\  {\rm and}\  b_n(x)$ are
formal power
series
with complex coefficients. The approach to non-critical string theory based on
the
consideration of matrix models [1]-[3] led to the problem of description of all
pairs
$P,Q$ satisfying $[P,Q]=1$ (see [4]). The equation $[P,Q]=1$
where $P$ and $Q$ are differential operators is known therefore
as string equation. We will assume that the orders $p$ and $h$ of differential
operators $P$
and $Q$ have no common divisors and that $Q$ is monic (i.e. the leading
coefficient $b_h
(x)$ is equal to $1$).
The set of all pairs $(P,Q)$ obeying these conditions will be denoted by ${\cal
A}_{p,h}$.

Every monic operator
$Q=\sum _{r=1}^h b_r(x)\partial^r$ can be normalized by means of transformation
$Q\rightarrow \tilde{Q}=e^{\gamma}
Qe^{-\gamma}$(i.e. one can make $b_{h-1}=0$).If $(P,Q)\in {\cal A}_{p,h}$  then
$(\tilde{P}, \tilde{Q})$ where
 $\tilde{P}=e^{\gamma}Pe^{-\gamma}, \ \
\tilde {Q}=e^{\gamma}Qe^{-\gamma}$ belongs to ${\cal A}_{p,h}$ too. Therefore
the study
of the space ${\cal A}_{p,h}$
can be reduced to the study of the space  ${\cal Q}_{p,h}$ consisting of pairs
$(P,Q)\in
{\cal A}_{p,h}$ where $Q$ is  normalized.

    One of our aims is to describe the set ${\cal Q}_{p,h}$. We give the
following
description. Let us denote by ${\cal M}_{p,h}$ the space of polynomial $h\times
h$
matrices $P=(P_{ij}(u))$ satisfying  $p=\max_{1\leq j\leq h}(j-i+h \deg
P_{ij})$ for every $i$.
(Here $i,j=1,\ldots,h, \deg P_{ij}$ denotes the degree of the polynomial
$P_{ij} (u)$).

  The group ${\cal T}_h$ of invertible triangular $h\times h$ matrices
$T=(t_{ij}),\ t_{ij}=0$
for $i<j$, acts on
${\cal M}_{p,h}$ by the formula $P\rightarrow TPT^{-1}$. We construct
one-to-one
correspondence between
${\cal Q}_{p,h}$ and $h$-fold
covering of the quotient space ${\cal M}_{p,h}/{\cal T}_h$. This construction
is based on
some theorems that have other interesting applications too. In particular these
theorems
can be used to give very simple and natural proof of some
results of [8], [9], [10].

  The description of the set of solutions to the string equation can be
formulated also in a
different way. Let us consider a set of pairs $(P,Q)$ of commuting \ddo s:
$[P,Q]=0$ where
$Q$ is a monic normalized operator of order $h,\
  P$ is an operator of order $p$ where $p$ and $h$ are relatively prime. The
set of all such
pairs will be denoted by ${\cal N}_{p,h}$. We will prove there exists
one-to-one
correspondence between ${\cal N}_{p,h}$ and $h$-fold covering of ${\cal
M}_{p,h}/{\cal
T}_h$ and therefore one-to-one correspondence between ${\cal Q}_{p,h} $ and
${\cal N}_{p,h} $. Using well known description of ${\cal N}_{p,h} $ in the
terms of moduli
spaces of algebraic curves we obtain a similar description of ${\cal Q}_{p,h}
$.

   Let us define the space ${\cal Q}^{\prime}_{p,h}$ as the space of such
operators $Q$
that one can find $P$ satisfying
   $(P,Q)\in{\cal Q}_{p,h}$. In the case when  $(P,Q)\in{\cal Q}_{p,h}$ and
$(P^{\prime},Q)\in{\cal Q}_{p,h}$ one can prove that $P^{\prime}-P=\sum
\sigma_kQ^k,\ \
0\leq hk<p.$ It follows from this fact that the description of ${\cal Q}_{p,h}$
given above
leads to similar description of ${\cal Q}^{\prime}_{p,h}$. Namely one has to
replace ${\cal
M}_{p,h}$ by its subset ${\cal M}^{\prime}_{p,h}$ singled out by the condition
$TrP=0$.

   Our results can be generalized to the supersymmetric case; some
generalizations will be
sketched at the end of the paper.

   The standard approach to the analysis of the string equation is based on the
Sato theory
of $KP$-hierarchy [6] (The
   most appropriate for our aims exposition of this theory is given in [7].)
Our approach is
based on more geometrical part of the same theory connected with
infinite-dimensional \gr
\ $Gr$. Let us denote by $H$ the space of formal Laurent series $\sum a_nz^n$,
where
$a_n=0$ for $n\gg0$. The
   subspace of $H$ spanned by $z^n$ with $n\geq 0$ will be denoted by $H_+$. If
$V$ is a
linear subspace of $H$ we will write $V\in Gr^{(0)}$ if the natural projection
$\pi_+$ of $V$
into $H_+$ is an isomorphism of $V$ and $H_+$. (In other words
$V\in Gr^{(0)}$ iff $V$ has a basis of the form $\varphi _n=z^n+$lower order
terms, $n\geq
0$). The set $Gr^{(0)}$ can be interpreted as the big cell of the \gr \
$Gr$ but this interpretation as well as the definition of $Gr$ is not necessary
for
our aims.

One can prove the following theorem:
There exists one-to-one correspondence between the set ${\cal Q}^{\prime}_{p,h}
$ and
the set $\tilde{{\cal Q}}_{p,h}$ of the elements $V\in Gr^{(0)}$ satisfying
\begin {equation}
z^hV\subset V,\ \ AV\subset V
\end {equation}
for some operator $A$ having the form
\begin {equation}
A={d \over d(z^h)}+\sum_{k=-h} \alpha_kz^k,\ \ \alpha_k=0\ \ {\rm for}\ k>p,\
\alpha_p\not=
0.
\end {equation}

The proof of this theorem  was worked out in discussions with Jeff
Rabin.\footnote{This
theorem is closely related to the results of [11--13]. After completion of this
paper I was
informed that another direct proof of this theorem was given by M.~Fukuma,
H.~Kawai,
and R.~Nakayama, preprint UT-582, May 1991.} It is based on the notion of \ps \
($\Psi$DO). By definition an expression of the form $S=\sum_{i\leq
k}s_i(x)\partial ^i$ is
called a \ps \ of order $k$. This operator is monic if $s_k(x)=1$, it is
normalized if
$s_{k-1}(x)=0$. (Here $s_i(x)$ are formal power series). The \ps s form an
algebra ${\cal
P}$; monic \ps s of order $0$ form a group; we will denote this group by ${\cal
G}$. One
can define an action of the algebra ${\cal P}$ on the space $H$ assuming that
$x^m\partial^n\in {\cal P}$ transforms $\varphi\in H$ into $({1\over i}{d\over
dz})^m
(iz)^n\varphi\in H$. It is well known that every $V\in Gr^{(0)}$ can be
represented in the
form $V=SH_+$ where $S\in {\cal G}$. If $V\in \tilde{{\cal Q}}_{p,h}$ then (2)
can be
interpreted as invariance of $V$ with respect to the action of $\Psi$DO's
$\partial ^h$ and
$\hat {A}$ where
\begin {equation}
\hat {A}=h^{-1}i^h\partial ^{1-h}\cdot x+\sum \alpha_k({1\over i}\partial)^k,\
\ \alpha_k=0\
{\rm for}\  k>p,\ \alpha_p\not= 0
\end {equation}

 This means that the operators $Q=S^{-1}\partial ^hS$ and $P=S^{-1}{\hat A}S$
transform
$H_+$ into itself. It is well known that $\Psi$DO transforming $H_+$ into
itself is a \ddo \
and therefore our construction gives an element of  ${\cal Q}^{\prime}_{p,h}$
for every  h

  Conversely if $Q$ is a monic normalized \ddo \ of order $h$ one can find such
an
operator $S\in {\cal G}$ that $\partial ^h=SQS^{-1}$ in the algebra ${\cal A}$.
If $P$ is a
\ddo \ of order $p$ and satisfies $[P,Q]=1$ then $\hat {A}=SPS^{-1}$ is a \ps \
obeying
$[\hat {A},\partial^h]=1$ and therefore can be represented in the form (4).
There is a freedom in the choice of the operator $S$ satisfying $\partial
^h=SQS^{-1}$.
Namely, this operator can be replaced by $RS$ where $R=\sum _{i=0} c_i\partial
^{-i}$ is a
monic $\Psi$DO with constant coefficients. One can use this freedom to obtain
an operator
$\hat{ A}$ obeying $\alpha_i=0$ for $i<-h$. It is easy to check that the space
$V=SH_+$
obtained by this choice of $S$ belongs to
$ \tilde{{\cal Q}}_{p,h}$. This assertion permit us to identify   ${\cal
Q}^{\prime}_{p,h}$
and  $\tilde{{\cal Q}}_{p,h}$.

 It is clear that in the definition of $\tilde{{\cal Q}}_{p,h}$ we can assume
without loss of
generality that in (3) $\alpha_k=0$ for $k=0,h,2h,\ldots$. (If $AV\subset V,\ \
z^hV\subset
V$ and $A^{\prime}=A-\sum\alpha_{ih}z^{ih}$ then $A^{\prime}V \subset V$.) We
will
impose this condition on the operator (3); then one can prove that for $V\in
\tilde{{\cal
Q}}_{p,h}$ there exists only one operator $A$ of the form (3) satisfying
$AV\subset V$.
Note that $\alpha _{-h}\not= 0$; we will prove that $\alpha_{-h}=-(h-1)/2$. The
numbers
$\alpha_i=\alpha_i(V)$ depend on the choice of $V$ if $i$ is not divisible by
$h$.

  We will use the theorem above to prove some general statements about the
solutions to
the string equation and to give an effective description of
  ${\cal Q}_{p,h}$.

   First of all, it follows immediately from the identification of
      ${\cal Q}^{\prime}_{p,h}$ and   $\tilde{{\cal Q}}_{p,h}$ that the string
equation is
invariant with respect to KP-flow. The $r$-th  KP-flow on Gr is described by
the formula
\begin {equation}
{d\over dt_r}V=z^rV,\ \ V\in Gr
\end {equation}
 It is clear that  $ \tilde{\cal Q}_h = \bigcup_p \tilde{{\cal Q}}_{p,h}$ is
invariant with respect
to the KP-flow on Gr. Really if  $V\in \tilde{{\cal Q}}_{p,h}$ satisfies
$AV\subset V$ then
the space $V(t_1,t_2,\ldots)$ satisfies
$A(t_1,t_2,\ldots)V(t_1,t_2,\ldots)\subset
V(t_1,t_2,\ldots)$ where $A(t_1,t_2,\ldots)$ has the form (3) and can be
determined by the
equations
\begin {equation}
{\partial A\over\partial t_r}=[z^r,A]=-{r\over h}z^{-h+r}.
\end {equation}

It follows from (6) that the coefficients $\alpha_i$ in the representation (3)
for $A$ satisfy
\begin {equation}
{\partial \over\partial t_r}\alpha_i=-{r\over h}\delta _{i,r-h}.
\end {equation}

 We see that
\begin {equation}
\alpha_i(t_1,t_2,\ldots)=\alpha_i-{i+h\over h}t_{i+h}.
\end {equation}
The correspondence $S\rightarrow V=S^{-1}H_+$ between the group ${\cal G}$ of
monic
zeroth order $\Psi$DO and $Gr^{(0)}$ gives KP-flow on ${\cal G}$:
\begin {equation}
{dS\over dt_r}=-(S\partial^r S^{-1})_-S
\end {equation}
where $B_-=B-B_+,B_+$ denotes the differential part of $\Psi$DO $B$. For
$L=S^{-1}\partial S$ we obtain the standard equation of KP-hierarchy
\begin {equation}
{dL\over dt_r}=[(L^r)_+,L].
\end {equation}

If $z^hV\subset V$ then the operator $Q=
S^{-1}\partial ^h S$ transforms $H_+$ into itself and therefore $Q$ is a \ddo .
The correspondence between $\tilde {{\cal Q}}_{p,h}$ and  ${\cal
Q}^{\prime}_{p,h}$
constructed above shows that ${\cal Q}^{\prime}_{p,h}$ is invariant with
respect to the
$h$-reduced KP-hierarchy:
\begin {equation}
{dQ\over dt_r}=[(Q^{r\over h})_+,Q], \; r \leq p+h, \: r \neq kh,
\end {equation}

This fact is proved in [10]; the relation (8) agrees with the results of [10].
The partition
function of 2D-gravity is connected with the $\tau$-function
$\tau _V (t_1,t_2,\ldots)$  corresponding to the element $V\in \tilde {{\cal
Q}}_{p,h}$. The
relation (2) means that this $\tau$-function does not depend on
$t_h,t_{2h},\ldots$ and satisfies
\begin {equation}
L_k\tau =0,\ \ k\geq -1,
\end {equation}
where
\begin {equation}
L_k={1\over h}\sum_{2\alpha <kh}J_{\alpha }J_{kh-\alpha}+{1\over 2h}J_{kh/2}^2+
{h^2-1\over 24h}\delta _{k0}+\sum a_{\alpha}J_{(k+1)h+\alpha},
\end {equation}
$J_{\alpha}=\partial/\partial t_{\alpha}$ for $\alpha>0,\  J_{\alpha}$ is a \ml
\ by $\alpha
t_{-\alpha}$ for $\alpha <0$. This assertion is proved in [11], [12], but used
in different
direction. Namely the equation (13) proved in [8], [9] was used in [11], [12]
to describe the
element of Sato \gr \  corresponding  to the $\tau$-function arising in
2D-gravity. Now we
see that (12) as well as more general $W$-constraints on the $\tau$-function is
a
consequence of the following remark: if $V$ satisfies (2) then
\begin {equation}
z^{kh}A^lV\subset V,\ \ k\geq 0,\ \ l\geq 0
\end {equation}
(The constraints (12) correspond to $l=1$. See [13] for the analysis of $W$-
constraints.)

  Now we will turn to the effective description of the set $\tilde {{\cal
Q}}_{p,h}$. Let us
consider an element $V\in  \tilde {{\cal Q}}_{p,h}$ and the vectors

  $\varphi_1,\ldots,\varphi _h\in V$ having the form

\begin {equation}
\varphi_k=c_kz^{k-1}+{\rm lower\  order\ terms},\ \ c_k\not= 0.
\end {equation}
It follows from $z^hV\subset V$ that the vector  $z^{sh}\varphi_i,\ \
i=1,\ldots,h;\ \
s=0,1,\ldots,$ belong to $V$. Taking into account that $AV\subset V$ we see
that
\begin {equation}
A\varphi _i=\sum_jP_{ij}(z^h)\varphi_j,\ \ 1\leq i,\ j\leq h,
\end {equation}
where $P_{ij}$ are polynomials. We can simplify the equations (16) by the
substitution
$\psi_i=\rho \varphi_i$ where $\rho(z)$ is a function satisfying $(d/dz^h)\cdot
\hat{\rho}=\hat{\rho} A$ (here $\hat{\rho}$ denotes the operator of \ml \ by
$\rho$ ). The
substitution  leads to the equations of the form (16) with $A$ replaced by
$d/dz^h$. Then
we make the substitution $\tau =z^{-i+1}\varphi _i$ and obtain
\begin {equation}
{d\tau _i\over dz}=\sum_j B_{ij}(z)\tau_j(z)
\end {equation}
where
\begin {equation}
B_{ij}(z)=hz^{j-i+h-1}P_{ij}(z^h)-(i-1)z^{-1}\delta_{ij}
\end {equation}
 The functions (or,more precisely, formal expressions) $\tau_i(z)$ have the
form
\begin {equation}
\tau_i(z)=\rho(z)\cdot (t_{i0}+t_{i1}z^{-1}+t_{i2}z^{-2}+\ldots),\ \
t_{i0}\not= 0
\end {equation}

where
$$\rho (z)=\exp R(z),\ \ {dR\over d(z^h)}=\sum _{k=-h}\alpha_kz^k $$
\begin {equation}
R(z)=h\alpha_{-h}\ln z+h\sum_{k=-h+1}{\alpha_k\over k+h}z^{k+h}
\end {equation}

 The equations for $\psi_i$ are invariant with respect to the transformation
$z\rightarrow
\epsilon z$,where $\epsilon ^h=1,\  \epsilon ^k\not= 1$ for $0<k<h$. Therefore
the system
(17) is invariant by the substitution $z\rightarrow\epsilon z,\
\tau_k\rightarrow\epsilon
^{-k}\tau_k$. If we know a solution $\tau(z)=(\tau_1(z),\ldots,\tau_h(z)) $ of
the system (17)
this invariance permits us to find a family of solutions
\begin {equation}
\tau^{(\lambda)}(z)=(\epsilon^{-\lambda}\tau_1(\epsilon^{\lambda}z),
\ldots,\epsilon^{-h\lambda}\tau_h(\epsilon^{\lambda}z)),
\ \ \lambda=0,1,\ldots,h-1.
\end {equation}

 We assumed that $\alpha_p z^p$ is the leading term in $\sum \alpha_k z^k$ and
that $p$
and $h$ have no common divisors, therefore the solutions (21) are linearly
independent.
Let us consider the wronskian
\begin {equation}
W(z)=\det(\epsilon^{-k\lambda}\tau_k(\epsilon^{\lambda}z)).
\end {equation}

We obtain by means of (19), (20), that
\begin {equation}
W(z)=\exp
(\sum_{\alpha=0}^{h-1}R(\epsilon^{\alpha}z))\sigma(z)=\exp(h\alpha_{-h}\ln
z)\sigma(z)
\end {equation}
where $\sigma(z)=\sigma _0+\sigma_1 z^{-1}+\sigma_2 z^{-2}+\ldots$. We used the
equation
\begin {equation}
\sum _{\lambda=0}^{h-1}(\epsilon^{\lambda}z)^k=0\ \ {\rm if}\ k\ {\rm is \ not\
divisible\ by}\
h
\end {equation}
and our assumption $\alpha _r=0$ for $r=0,h,2h\ldots$. From the other side it
is well known
that
\begin {equation}
W(z)=W(0)\exp \int _0^z TrB(z) dz.
\end {equation}

Taking into account that
\begin {equation}
TrB(z)=hz^{h-1}TrP(z^h)-{h(h-1)\over 2}z^{-1}
\end {equation}
we obtain
\begin {equation}
W(z)=W(0)\exp(\int _0^{z^h} Tr P(u)du-{h(h-1)\over 2}\ln z).
\end {equation}

Comparing (27) and (23) we conclude that
\begin {equation}
TrP=\sum _i P_{ii}(u)=0,
\end {equation}
\begin {equation}
\alpha_{-h}={h-1\over 2}.
\end {equation}
In such a way for every element $V\in \tilde {{\cal Q}}_{p,h}$ we constructed a
polynomial
matrix $P_{ij}$.
The construction was based on the choice of  vectors $\varphi_1,\ldots
,\varphi_h\in V$
satisfying (15). Of course
we can replace these vectors by $\tilde{\varphi}_i=\sum t_{ij}\varphi_j$ where
$T=(t_{ij})$
is an invertible constant triangular matrix $(t_{ij}=0$ for $i<j$). This change
of vectors
$\varphi_1,\ldots ,\varphi_h$ induces the change of the matrix $P=(P_{ij})$ by
the formula:
\begin {equation}
\tilde {P} =TPT^{-1}.
\end {equation}

The matrix $P$ corresponding to an element $V\in\tilde {{\cal Q}}_{p,h}$
satisfies (28). The
degree of the leading term $\tilde {B}$ of the matrix  $B=(B_{ij})$ is equal to
$s+h-1$ where
$$s=\max _{1\leq i,j\leq h}(j-i+h\deg P_{ij})$$
The matrix $\tilde {B}$ has the form
$$\tilde {B}_{ij}=\beta_i\delta_{i,j-s}z^{s+h-1}$$
(We consider here the indices $i,j$ as residues mod $h$.)
The leading terms $t_{i0}$ in the expressions (19) for $\tau_i(z)$ satisfy the
equation

$$\sum _j \beta_i\delta_{i,j-s}t_{j,0}=\alpha_st_{i,0}.$$
It follows immediately from the condition $t_{i,0}\not= 0$ that $\alpha_s\not=
0$ and
$\beta_i\not= 0$ for every $i=1,\ldots ,h$.
It is easy to check that $\alpha_s z^s$ is the leading term in the sum $\sum
\alpha_iz^i$
and therefore $s=p$ and $h$ have no common divisors. The matrix $\tilde {B}$
has exactly
one non-zero entry in every row (this follows from $\beta_i\not= 0$). The
information about
the matrix $\tilde {B}$ obtained above is equivalent to the following
statement:
\begin {equation}
p=\max_{1\leq j\leq h} (j-i+h\deg P_{ij})\ {\rm for \ every}\ i,\ 1\leq i\leq
h.
\end {equation}
One can prove that the conditions (28) and (31) are sufficient to assert that
the matrix
$P_{ij}$ corresponds to an element  $V\in\tilde {{\cal Q}}_{p,h}$. The proof
repeats the
arguments of [12]. We have to prove that (17) has a formal solution of the form
(19) where
the numbers $\alpha_k$ in the expression (20) satisfy $\alpha_k=0$ for
$k=0,h,2h,\ldots.$
Then it is easy to check that the formal powers series $z^{sh}\varphi_i,\
s=0,1,\ldots,h$,
where $\varphi_i=z^{i-1}\rho^{-1}\tau_i$, span a space $V\in\tilde {{\cal
Q}}_{p,h}$. The
construction of solutions to (17) is based on the remark that the \ei \ of the
leading term of
the matrix $B(z)$ are distinct. This follows from the condition (31). Really,
the invariance of
(17) with respect to the transformation $z\rightarrow \epsilon z,\ \
\tau_k\rightarrow
\epsilon^{-k}\tau_k$ leads to the following property of $B(z)$:
\begin {equation}
\epsilon B(\epsilon z)=\Pi B(z)\Pi^{-1}
\end {equation}
where $\Pi$ is a matrix with the entries
$\pi_{ik}=\epsilon^{-(i-1)}\delta_{ik}$. If the leading
term $\tilde {B}$ of
$B(z)$ is $B_{m-1}z^{m-1}$ we obtain from (32) that

\begin {equation}
\epsilon ^mB_{m-1}=\Pi B_{m-1}\Pi ^{-1}
\end {equation}

If $B_{m-1}l=\lambda l$ then $\Pi^kl$ is an eigenvector of $B_{m-1}$ with an
eigenvalue
$\epsilon^{-km}\lambda$. We obtain $h$ distinct \ei \ of $B_{m-1}$ in this way.
(We use
that $m=s+h$ and $h$ are relatively prime.)

  Using perturbation theory we can find formal expansions (Laurent series) for
\ei \
$\lambda_1(z),\ldots,\lambda_h(z)$ of $B(z)$. It follows from (33) that we can
label these
\ei \ in such a way that

\begin {equation}
\epsilon \lambda_k(\epsilon z)=\lambda_{k+1}(z).
\end {equation}

If $\lambda_1=\sum \nu_i z^i$  we have

\begin {equation}
TrB(z)=\sum \lambda_k(z)=\sum\epsilon^{k-1}\lambda_1(\epsilon^{k-1}z)=\sum
\epsilon
^{(k-1)(i+1)}\nu_iz^i=\sum h\nu_{nh-1}z^{nh-1}.
\end {equation}

Taking into account (26) and (28) we get

$$TrB(z)=-{h(h-1)\over 2}z^{-1}$$
and therefore $\nu_{nh-1}=0$ for $n>0$.

   In the case when the \ei \ of the leading term of $B(z)$ are distinct one
can construct
formal

    solutions to (17) in the form (19) where $\rho(z)=\int \lambda_k(z) dz$.
The condition (31)
gives that $t_{k0}\not= 0$. It follows from $\nu_{nh-1}=0$ that for these
solutions
$\alpha_{kh}=0$ for $k=0,1,\ldots$ in the formula (20) and therefore using
these solutions
one can construct elements of $\tilde {{\cal Q}}_{p,h}$.
    In such a way for every polynomial $h\times h$ matrix $P=(P_{ij})$
satisfying (28) and
(31) we can construct $h$

 elements of  $\tilde {{\cal Q}}_{p,h}$ labeled by \ei \ of the leading term
$\tilde {B}$ of the
matrix $B$. Two matrices $P$ and $P^{\prime}$ correspond to the same element of
$\tilde
{{\cal Q}}_{p,h}$ in the case when they are connected by the formula
$P^{\prime}=TPT^{-1}$ where $T$ is a triangular matrix (and only in this case).
Using the
notation ${\cal M}^{\prime}_{p,h}$ for the space $h\times h$ matrices
satisfying (28) and
(31) we can identify $\tilde {{\cal Q}}_{p,h}$, with $h$-fold covering of
${\cal
M}^{\prime}_{p,h}/{\cal T}_h$, where ${\cal T}_h$ is the group of constant
triangular
matrices.

    Thus, we gave a description of $\tilde{\cal Q}_{p,h}$ and, therefore, a
description of the
moduli space ${\cal Q}_{p,h}$. Our next aim is to study the connection between
${\cal
Q}_{p,h}$ and the moduli space of Riemann surfaces (of algebraic curves). We
will begin
from the study of pairs of commuting \ddo s:$[P,Q]=0$, where $Q$ is a monic
normalized
\ddo . Let us denote by ${\cal N}_{p,h}$ the space of pairs of commuting \ddo s
$P,Q$
where ord$P=p$, ord$Q=h$, and $p,h$ are relatively prime. We will prove that
there exists
one-to-one correspondence between ${\cal N}_{p,h}$ and ${\cal Q}_{p,h}$. It
will be more
convenient for us to work with the space ${\cal N}^{\prime}_{p,h}$ obtained
from ${\cal
N}_{p,h}$ by means of identification $(P,Q)\sim (P+\sum\alpha _k Q^k, Q)$ and
construct
one-to-one correspondence between ${\cal N}^{\prime}_{p,h}$ and ${\cal
Q}^{\prime}_{p,h}$. The first step is the well known construction connecting
commutative
rings of \ddo s with rings of functions. Let us consider $\Psi$DO $S$
satisfying
$\partial^h=SQS^{-1}$ in the algebra ${\cal P}$. (We have mentioned that such
an operator
exists and is defined up to  \ml \ from the left by monic $\Psi$DO $\sum
a_i\partial ^{-i}$
with constant coefficients $a_i$.) Then the space $V=SH_+\in Gr^{(0)}$ is
invariant with
respect to the action of $\Psi$DO's $\partial ^h$ and $SPS^{-1}$. The action of
$\partial
^h$ in $H$ can be interpreted as \ml \ by $z^h$  and the action of $SPS^{-1}$
can be
considered as \ml \  by a series $b(z)=\sum b_iz^i\in H$ (because $SPS^{-1}$
commutes
with $\partial ^h$). The degree of the leading term $b_pz^p$ of $b(z)$ is equal
to the order
of the operator $P$ and therefore it is relatively prime to $h$. If we replace
the pair $(P,Q)$
by ($P+\sum \alpha_kQ^k,Q$) the series $b(z)$  changes to
$b(z)=\sum\alpha_kz^{hk}$
and therefore we can make $b_k=0$ for $k=0,h,2h,\ldots $.

   Let us define by $\hat{{\cal N}}_{p,h}$ the set of all pairs $(b,V)$ where
$V\in Gr^{(0)},\
b(z)=\sum b_i z^i\in H,\  b_i=0$ for $i=0,h,2h,\ldots $ the degree $p$ of the
leading term of
$b(z)$ is relatively prime to $h,\  z^hV\subset V,\  b(z)V\subset V$. The group
of monic
$\Psi$DO's with constant coefficients acts in $ {\cal N}_{p,h}$ (this action
can be realized
by means of multiplication by formal Laurent series
$1+c_1z^{-1}+c_2z^{-2}+\ldots).$ The
space of orbits of this action will be denoted by $\tilde {{\cal N}}_{p,h}$.
The  construction
above gives one-to-one correspondence between $\tilde{{\cal N}}_{p,h}$ and
${\cal
N}_{p,h}$. To prove this fact one has to construct the inverse map, taking into
account that
for every $V\in Gr^{(0)}$ there exists a monic zeroth order $\Psi$DO $S$
satisfying
$V=SH_+$. (Transforming $\Psi$DO's corresponding to the \ml \  by $b(z)$ with
the help of
operator $S$ we obtain commuting $\Psi$DO's. These $\Psi$DO's preserve $H_+$
and
therefore can be considered as \ddo s.)

    Our next problem is to describe the set $\tilde{{\cal N}}_{p,h}$. Let us
take a pair
$(b,V)\in \hat{{\cal N}}_{p,h}$ and the elements
$\varphi_1,\varphi_2,\ldots,\varphi_h\in
V$ satisfying $\varphi_i=z^{i-1}+$lower order terms. It is clear that

\begin {equation}
b(z)\varphi_i=\sum_j P_{ij}(z^h)\varphi_j
\end {equation}

where $P_{ij}$ are polynomials with respect to $z^h$. Introducing the notation
$\sigma
_i=z^{1-i}\varphi_i$ we can write (36) in the form
\begin {equation}
b(z)\sigma_i=\sum_j C_{ij}\sigma_j
\end {equation}

where

\begin {equation}
C_{ij}=z^{j-i}P_{ij}(z^h).
\end {equation}

  The matrix $C(z)$ is related to the matrix $B(z)$ by the formula
$B(z)=hz^{h-1}C(z)-(i-1)z^{-1}\delta_{ij}$ and has similar symmetry properties.
Namely,

\begin {equation}
C(\epsilon z)=\Pi C(z)\Pi ^{-1}.
\end {equation}

  We see from (37) that $(\sigma_1,\ldots ,\sigma_h)$ is a formal eigenvector
of $C(z)$ with
the eigenvalue $b(z)$. It follows from (39) that $C$ has also  \ei \
$\lambda(z)=b(\epsilon
^k z)=\sum b_n\epsilon ^{kn}z^n$ corresponding to the eigenfunctions
$\Pi^{-k}\sigma(\epsilon ^k z)$.

   Using (24) we get

\begin {equation}
TrC(z)=\sum_{1\leq k\leq h}\lambda_k(z)=\sum_{1\leq k\leq h}\sum_n
b_n\epsilon^{kn
}z^n=\sum b_{ih}z^{ih}
\end {equation}

Noting that $b_{ih}=0$ for $i\geq 0$ and that $TrC(z)$ is a polynomial we
obtain

\begin {equation}
TrC(z)=TrP(z)=0
\end {equation}

We see that the matrix $P(z)$ satisfies (28). Slight modification of the
arguments above
shows also that $P(z)$ satisfies (31). Taking into account that the vectors
$\varphi_1,\ldots
,\varphi_h$ are determined up to triangular linear transformation we see that
assigning the
matrix $P=(P_{ij})$ to the pair $(b,V)\in \hat{{\cal N}}_{p,h}$ we obtain a map
from
$\hat{{\cal N}}_{p,h}$ into ${\cal M}_{p,h}/{\cal T}_h$. It is easy to check
that this map
induces

a map from $\tilde{{\cal N}}_{p,h}$ into ${\cal M}^{\prime}_{p,h}/{\cal T}_h$.
Conversely, if
the matrix $P_{ij}$ satisfies (28),(31) we can assert that the leading term of
the matrix
$C(z)$ has distinct \ei \  and therefore one can construct \ei \  $b(z)$ and
eigenvectors
$\sigma_1,\ldots ,\sigma_h$ of the matrix $C(z)$ as formal power series using
standard
methods of perturbation theory. Knowing these \ei \  and eigenvectors we can
reconstruct
a point of $\tilde{{\cal N}}_{p,h}$ corresponding to the matrix $P_{ij}$. (For
every matrix
$P_{ij}$ we can construct $h$  \ei \ labeled by  the \ei \  of the leading
term of $C(z)$. The
eigenvectors are specified up to a factor of the form
$1+a_1z^{-1}+a_2z^{-2}+\ldots $,
therefore we obtain a point in $\tilde{{\cal N}}_{p,h}$ instead of a point in
$\hat {{\cal
N}}_{p,h}$.) In such a way we obtain an identification of ${\cal N}_{p,h}$ with
$h$-fold
covering of ${\cal M}^{\prime}_{p,h}/{\cal T}_h$ and therefore an
identification of ${\cal
N}^{\prime}_{p,h}=\tilde{{\cal N}}_{p,h}$ with ${\cal
Q}^{\prime}_{p,h}=\tilde{{\cal
Q}}_{p,h}$.

 In such a way we described the set ${\cal Q}_{p,h}$ of solutions to the string
equation
$[P,Q]=1$ and constructed one-to-one correspondence between this set and the
set
${\cal N}_{p,h}$ of pairs of commuting  \ddo s. It is well known that the pairs
of commuting
\ddo s are closely related to algebraic curves. Namely if $(P,Q)\in {\cal
N}_{p,h}$ the
operators $P$ and $Q$ are connected by polynomial relation $F(P,Q)=0$. This
relation
determines an algebraic curve. Conversely let us take an  algebraic curve $K$
and a line
bundle $L$ over $K$. If $\deg L={\rm genus}(K)-1$ one can construct pairs
$(b,V)\in
\hat{{\cal N}}_{p,h}$ by means of Krichever construction. (For detailed
explanation of the
connection between the set of commuting pairs of \ddo s and the \ms \  of
algebraic curves
see for example [7]).

  It is worthy to mention that it was suggested to consider pairs of \ddo s
satisfying
$[P,Q]=1$ as quantum analogs of Riemann surfaces (see e.g. [5]). One can
interprete
${\cal Q}_{p,h}$ as \ms \  of quantum Riemann surfaces. However more natural
definition
of the \ms \ of quantum Riemann surfaces requires identifications  $(P,Q)\sim
(P+\sum
\sigma_iQ^i,Q)$ and $(P,Q)\sim (Q,P)$. If only the first of these
identifications is taken into
account then the  \ms \ coincides with ${\cal Q}^{\prime}_{p,h}=\tilde{{\cal
Q}}_{p,h}$.

   It is important to stress that the consideration above can be generalized to
the
supersymmetric case by means of results of [16], [17]. There are different
superanalogs of
the string equation (see [15] for detailed analysis). Here we will study only
the equation
$[P, Q]=1$ where $P$ and $Q$ are even super\ddo s, assuming that the operator
$Q$ is
monic and normalized. In other words $P=\sum a_k(x,\xi )\partial ^k+\sum
\alpha_k(x,\xi
)\partial ^k\partial _{\xi},\ \ Q=\sum b_k(x,\xi )\partial ^k+\sum \beta
_k(x,\xi )\partial
^k\partial_{\xi},\ \ b_h(x,\xi)=1,\ \ b_k(x,\xi)=0$ for $k>h$ and $k=h-1,\ \
b_k(x,\xi)=0$ for
$k\geq h-1 $. (Here $x$ is an even variable, $\xi$ is an odd variable,
$\partial =\partial
/\partial x,\  \partial_{\xi}$ denotes the left derivative with respect to
$\xi,\ \ a_k(x,\xi)$ and
$ b_k(x,\xi)$ are even formal power series and $\alpha_k(x,\xi),\
\beta_k(x,\xi)$ are odd
formal power series.) The set of all pairs of even  super\ddo s satisfying the
conditions
above will be denoted by ${\cal Q}_h$.

   The superanalog of the space $H$ is the space of all formal Laurent series
$\sum a_n
z^n+\sum \alpha _nz^n\theta $ where $a_n=\alpha _n=0$ for $n\gg 0,\ z$ is even,
$\theta$
is odd. The superanalog  of $Gr^{(0)}$ is the set of linear subspaces of $H$
having a basis
$\varphi_1,\ \hat {\varphi}_1,\ \varphi_2,\ \hat{\varphi}_2,\ldots,$ where
$$
\varphi_i=z^{i-1}+\sum_{k>0}b_{ik}z^{-k}+\sum_{k<0}\beta_{ik}z^{-k}\theta$$
$$  \hat{ \varphi}_i=z^{i-1}\theta+\sum_{k>0}\gamma
_{ik}z^{-k}+\sum_{k>0}c_{ik}z^{-k}\theta.$$
We will use the same notations $H$ and $Gr^{(0)}$ for these objects. Let us
consider an
element $V\in Gr^{(0)}$ satisfying $z^hV\subset V,\ AV\subset V$, where

\begin {equation}
A=h^{-1}z^{1-h}{\partial\over\partial z}+\sum s_k z^k +\sum \sigma_k
z^k\theta+\sum\nu _k
z^k{\partial\over \partial \theta}+\sum v _k z^k\theta {\partial\over \partial
\theta},
\end {equation}
$s_k=\sigma_k=\nu_k=v_k=0$ for $i\gg 0$. The set of all pairs $(A,V)$
satisfying the
conditions will be denoted by $\tilde{{\cal Q}}_h$.

 Given a pair $(A,V)\in \tilde{{\cal Q}}_h$ one can construct an element of
${\cal Q}_h$; all
elements of ${\cal Q}_h$ can be obtained by means of this construction. The
proof of this
statement is based on the notion of super $\Psi$DO [14] and on the results of
[17]. Super
$\Psi$DO can be defined as an expression  $S=\sum c_i(x,\xi)\partial^i+\sum
\gamma_i(x,\xi)\partial^i\partial_{\xi}$ where $c_i(x,\xi)$ and
$\gamma_i(x,\xi)$ are formal
power series with respect to even variable $x$ and odd variable $\xi,\
c_i(x,\xi)=\gamma_i(x,\xi)=0$ for $i\gg 0$. Super $\Psi$DO's act on $H$ by
means of
Fourier transformation (i.e. $\partial$ and $\partial _{\xi}$ act as \ml \ by
$iz$ and by
$\theta$ correspondingly; $x$ and $\xi$ act as $i^{-1}\partial /\partial z $
and $\partial
/\partial \theta$). Monic zeroth order $\Psi$DO's constitute a group $G$ (i.e.
$S\in G$ iff
$c_i=\gamma_i=0$ for $i>0,\  c_0=1,\  \gamma_0=0$). It is proved in [17] that
for every
$V\in Gr^{(0)}$ there exists unique operator $S\in G$ obeying $V=S^{-1}H_+$.
(Here
$H_+$ denotes a subspace of $H$ spanned by $z^n,\  z^n\theta,\  n\geq 0$.)

  If $(A,V)\in \tilde{{\cal Q}}_h$ we can construct a pair $(P,Q)\in {\cal
Q}_h$ by the
formulas $P=S^{-1}\hat {A}S,\  Q=S^{-1}i^{-h}\partial ^hS$. (Here $S\in G,\
V=S^{-1}H_+,\
\hat{A} $ denotes the $\Psi$DO corresponding to the operator $A$.) Every pair
$(P,Q)\in
{\cal Q}_h$ can be obtained by means of this construction.

  The generalization of $KP$-hierarchy to the supersymmetric case was given by
Manin
and Radul [14] and by Mulase [17] and Rabin [18]. We will show that ${\cal
Q}_h$ is
invariant with respect to $KP$-flows both in Manin-Radul and Mulase-Rabin
sense. To
prove the invariance with respect to Manin-Radul  $KP$-flows we consider an
operator

  \begin {equation}
R(t_1,t_2,\ldots)=\sum_{i=1}^{\infty}t_iD^i
\end {equation}
where $D=\partial _{\xi}+\xi\partial,\  (t_1,t_3,\ldots)$ are odd parameters,

$(t_2,t_4,\ldots)$ are even parameters.
Let us introduce the notations

$$V(t_1,t_2,\ldots)=e^{R
(t_1,t_2,\ldots)}V,$$
$$\hat {A}(t_1,t_2,\ldots)=e^{R(t_1,t_2,\ldots)}\hat {A}
e^{-R(t_1,t_2,\ldots)}$$
It is easy to check that  $\Psi$DO $\hat {A} (t_1,t_2,\ldots)$ acts on $H$ as
an operator
$A(t_1,t_2,\ldots)$ having the form (42) with the coefficients

$$s_k(t_1,t_2,\ldots)=s_k-{h+k\over h}i^{h+k}t_{2(k+h)},
\sigma_k(t_1,t_2,\ldots)=\sigma_k-{h+k\over h}i^{h+k}t_{2(k+h)+1},$$
$$\nu_k(t_1,t_2,\ldots)=\nu_k -{h+k\over h}i^{h+k}t_{2(k+h)-1},
v_k(t_1,t_2,\ldots)=v_k$$

   Using this fact and the remark that $R$ commutes with $\partial ^h$ we
conclude that
$(A (t_1,t_2,\ldots),\  V(t_1,t_2,\ldots))\in \tilde {{\cal Q}}_h$. Let us
define
$S(t_1,t_2,\ldots)$ as an element of $G$ satisfying
$V(t_1,t_2,\ldots)=S^{-1}(t_1,t_2,\ldots)H_+$.In other words

$$S(t_1,t_2,\ldots)=Se^{-R (t_1,t_2,\ldots)}$$
where $S\in G$ obeys $V=S^{-1}H_+$.
 One can conclude from the results of [17] that $S (t_1,t_2,\ldots)$ obeys
equations
equivalent to the Manin-Radul $KP$-hierarchy. More precisely one can show that
the
operator

 $$L (t_1,t_2,\ldots)=S^{-1}(t_1,t_2,\ldots)DS(t_1,t_2,\ldots)$$
 satisfies Manin-Radul $KP$-hierarchy. The pair $(P(t_1,t_2,\ldots),
Q(t_1,t_2,\ldots))$
where

 $$P(t_1,t_2,\ldots)=S^{-1}(t_1,t_2,\ldots)\hat {A}S(t_1,t_2,\ldots),$$
 $$Q(t_1,t_2,\ldots)=S^{-1}(t_1,t_2,\ldots)i^{-h}\partial ^h
S(t_1,t_2,\ldots)=i^{-h}L^{2h}(t_1,t_2,\ldots)$$
belongs to ${\cal Q}_h$.

   In such a way we proved the invariance of the superstring equation with
respect to
Manin-Radul  $KP$-flows. (Note that this statement does not contradict to [10]
because our
definitions are slightly different: we do not assume that $P$ is a fractional
power of $Q$.) In
the case of Mulase-Rabin $KP$-hierarchy one can use the same arguments
replacing the
operator (43) by the operator

$$T(t_1,t_2,\ldots)=\sum t_{2m}\partial ^m+\sum
t_{2m+1}\partial^m\partial_{\xi}.$$
and using the fact that for $S\in G$ the operator
$S(t_1,t_2,\ldots)=Se^{-T(t_1,t_2,\ldots)}$
is

a solution to the Mulase-Rabin $KP$-hierarchy.

 We described the symmetries of supergeneralization of the string equation
corresponding
to the operators (43), (44). It is clear that every $\Psi$DO commuting with
$\partial^h$
generates such a symmetry. In other words(infinitesimal) symmetry
transformations of this
equation correspond to $\Psi$DO's

 $$\sum
s_k\partial^k+\sum\sigma_k\partial^k\partial_{\xi}+\sum\nu_k\xi\partial^k+\sum
v_k\xi\partial^k\partial_{\xi}.$$
Corresponding Lie superalgebra can be identified with the Lie algebra
$\tilde{gl}_{1|1}$
consisting of Laurent series $\sum g_nz^n$ where $g_n\in gl_{1|1}$. (Here
$gl_{1|1}$
denotes the Lie superalgebra consisting of linear transformations

of $(1|1)$-dimensional linear superspace.) These more general symmetry
transformations
are connected with Kac-van de Leur version of super $KP$-hierarchy.

   I am indebted to M. Mulase and J. Rabin for useful discussions.
   \vskip .1in
   \centerline {\bf References.}
   \vskip .1in
 1. Brezin, E.,  Kazakov, V., Phys. Lett. 236 B (1990) 144

 2. Douglas, M.,  Shenker, S., Nucl.Phys., B335(1990)635

 3. Gross, D.,  Migdal, A., Phys. Rev. Lett. 64 (1990) 127

 4. Douglas, M., Phys. Lett. 238 B (1990) 176

 5. Moore, G., Comm. Math. Phys., 133 (1990) 26

 6. Sato, N., RIMS Kokyuroku, 439 (1981) 30

 7. Mulase, M., International J. of Math., 1 (1990) 293

 8. Dijkgraaf, R.,  Verlinde, E. and H., IAS preprint (1990)

 9. Fukuma, M.,  Kawai, H.,  Nakayama, R., Univ. of Tokyo preprint, UT-562
(1990)

 10.Di Francesko, P.,  Kutasov, D., PUPT-1206 (1990)

 11. Kac, V.,  Schwarz, A, Phys. Lett. 257 B (1991) 329

 12. Schwarz, A., Mod. Phys. Lett. A6 (1991) 611

 13. Fukuma, M.,  Kawai, H.,  Nakayama, R., Univ. of Tokyo preprint, UT-572
(1990)

 14. Manin, Yu., Radul, A., Comm. Math. Phys. 98 (1985) 65

 15. Mulase, M., Schwarz, A., in preparation

 16. Mulase, M., Invent. Math. 92 (1988) 1

 17. Mulase, M., UC Davis preprint (1990) J. Diff. Geom. to appear

 18. Rabin, J, UCSD preprint (1990) Comm. Math. Phys. to appear

\end{document}